# Counting the electrons in a multiphoton ionization by elastic scattering of microwaves

A. Sharma, M.N. Slipchenko, M.N. Shneider, X. Wang, K.A. Rahman, A. Shashurin

Laser induced plasmas have found numerous applications including plasma-assisted combustion, combustion diagnostics, laser induced breakdown spectroscopy, light detection and ranging techniques (LIDAR), microwave guiding, reconfigurable plasma antennae etc. Multiphoton ionization (MPI) is a fundamental first step in high-energy laser-matter interaction and is important for understanding of the mechanism of plasma formation. With the discovery of MPI more than 50 years ago, there were numerous attempts to determine basic physical constants of this process in the direct experiments, namely photoionization rates and cross-sections of the MPI, however, no reliable data is available until today and spread in the literature values often reaches 2-3 orders of magnitude. This is due to inability to conduct absolute measurements of plasma electron numbers generated by MPI which leads to uncertainties and, sometimes, contradictions between the MPI cross-section values utilized by different researchers across the field. Here we report first direct measurement of absolute plasma electron numbers generated at MPI of air and subsequently we precisely determine ionization rate and cross-section of eight-photon ionization of oxygen molecule by 800 nm photons $\sigma_8=(3.32\pm 0.3)\cdot 10^{-130}$ $W^{-8}m^{16}s^{-1}$. Method is based on the absolute measurement of electron number created by MPI using elastic scattering of microwaves off the plasma volume in Rayleigh regime and establishes a general approach to directly measure and tabulate basic constants of the MPI process for various gases and photon energies.

**Introduction**

Since its first discovery in mid-1960s[1,2,3], laser-induced plasmas have found numerous application in the laboratories ranging from fundamental studies of nonequilibrium plasmas[4,5], soft ionization in mass spectroscopy[6], development of compact particle accelerators[7,8], X-ray and deep UV radiation sources[9,10] to diagnostic techniques such as laser induced breakdown spectroscopy and laser electronic excitation tagging[11,12,13]. In addition, the laser-induced plasma is related to studies of various nonlinear effects at beam propagation such as laser pulse filamentation, laser beam collapse, self-trapping, dispersion, modulation instability, pulse splitting etc[5,11,12,14,15].

The multiphoton ionization (MPI) is a key first step in all laser induced plasma. However, basic physical constants of the MPI process, namely photoionization rates and cross-sections have never been precisely determined in direct experiment. This refers to the fact that there are no diagnostic tools today to provide absolute measurements of total number of electrons in plasma volume $N_e$ or local plasma density distribution $n_e(\mathbf{r})$ generated by femtosecond laser pulse in relatively low intensity linear regime. For MPI of atmospheric air, plasma density has to be below $n_e \leq 10^{15}$-$10^{16}$ cm$^{-3}$ to ensure that contribution of plasma nonlinearities to the refraction index is negligible[5]. At the same time, the sensitivity of laser interferometry is limited to $n_e \geq 10^{16}$-$10^{17}$ cm$^{-3}$ due to the minimal measurable shifts of the interference fringes[16,17,18]. Number of semi-empirical methods for relative measurements of plasma density were proposed as well; however, all of them require absolute calibration based upon theoretically predicted values of plasma number density. Time-of-flight (TOF) mass spectrometer measurements of ion currents generated by laser-induced plasma filament have been conducted to measure photoionization rates[14,19,20]. The measurement relied on theoretical estimation of total number of electrons in focal zone in order to conduct absolute calibration of the system. Very recently, scattering of THz radiation from the laser-induced plasmas was proposed for spatially unresolved relative measurements of $n_e$[16,17]. Other measurement techniques were proposed recently based on measurements of capacitive response times of system including capacitor coupled with laser-induced plasma loaded inside[21,22,23]. These attempts to measure $n_e$ in laser-induced plasmas are characterized by various degrees of success and reliability of obtained data, but none of them provides ultimate solution for absolute plasma density measurements until today.

Analysis of various theoretical and semi-empirical approaches undertaken previously led to large variability of photoionization process constants available in literature and some of them being even controversial. For example, photoionization rates for $O_2$ reported by Mishima in ref. 24 are approximately 2 orders of magnitude higher than that reported by Talebpour in Ref.20. In addition, comparison of photoionization rates for $N_2$ and $O_2$ yields to 3 order of magnitude higher photoionization rates for $O_2$ compared to that of $N_2$ due to difference in ionization potentials (15.576 eV and 12.063 eV respectively)[24], while experimentally determined photoionization rates reported by Talebpour for $N_2$ and $O_2$ yields doubtful proximity, namely 1.5·10$^9$ s$^{-1}$ for $N_2$ and 3·10$^9$ s$^{-1}$ for $O_2$[20].

Therefore, photoionization rates and cross-sections of MPI process still remain unknown 50 year after the discovery. Large discrepancy in photoionization rates available in the literature causes sometimes contradictory conclusions and generally disadvantageous for theoretical modeling of wide class of problems involving laser-induced plasmas. In this work we propose direct experimental approach to measure total number of electrons created at the MPI of air and directly determine ionization rate and cross-section of MPI for oxygen molecule. Proposed approach has tremendous fundamental significance and great potential for applications since it paves the way to directly measure and tabulate basic constants of the MPI process for various gases and photon energies.

**Methodology of MPI cross-section determination**

Ionization of gas in laser-induced plasma is associated with multiphoton (MPI) and tunneling processes, which are two limiting cases of essentially same physical process of nonlinear photoionization. Choice of the governing mechanism is dictated by Keldysh parameter $\gamma$, defined as a ratio of laser frequency $\omega$ to tunneling frequency $\omega_t$ characterizing time of electron tunneling through the potential barrier: $\gamma = \frac{\omega}{\omega_t} = \frac{\omega\sqrt{2m\mathcal{E}_i}}{eE}$, where $E$- amplitude of incident electric field, $\mathcal{E}_i$- ionization potential, $m$ and $e$ are electron mass and charge respectively. In the case of low frequency limit (and/or large laser intensity) $\omega < \omega_t$, the electron has sufficient time to tunnel through the barrier and ionization is driven by the tunnel effect, while for high-frequency limit (and/or low laser intensity) $\omega > \omega_t$, electric field is varying faster than time required for tunneling and ionization is governed by the MPI process.

In our method, cross-section of the MPI is determined experimentally based on absolute measurement of total electron numbers ($N_e$) generated by a femtosecond laser pulse and precise measurements of the laser pulse characteristics. The experiments have been conducted at low laser intensities (<2.7·10$^{13}$ W/cm$^2$ as detailed below) in order to ensure pure linear operation regime when nonlinearities associated with plasma creation and optical Kerr effect are negligible (see below for details). In this case plasma formation due to MPI by the femtosecond laser is described by simple differential equation $\frac{\partial n_e}{\partial t} = \nu \cdot (n_0 - n_e)$, where $n_e$ - plasma density, $\nu = \sigma_m I^m$ – ionization rate, $\sigma_m$ – cross-section of $m$-photon ionization process with $m = \text{Int}\left(\frac{\mathcal{E}_i}{\hbar\omega}\right) + 1$, $I$– local instantaneous value of laser field intensity, and $n_0$ – background gas density, while other physical processes can be neglected on the extremely fast time scale of the laser pulse[1,25]. This equation can be easily integrated and plasma density $n_e$ created as result of action of femtosecond laser pulse can be found: $n_e = n_0\left(1 - e^{-\int \nu dt}\right)$. For the case of low ionization degree $n_e \ll n_0$ plasma density distribution immediately after the laser pulse can be written in the form:

$$n_e(\mathbf{r}) = n_0 \int \nu dt = \sigma_m n_0 \int I(\mathbf{r},t)^m dt \qquad (1)$$

where time integration is taken over the duration of the laser pulse at particular location $\mathbf{r}$. Total electron number $N_e$ generated by the laser pulse can be expressed by integrating Eq.(1) over the entire plasma volume:

$$N_e = \sigma_m n_0 \int \int I(\mathbf{r},t)^m dt dV \qquad (2)$$

Eq. (2) provides general expression which can be used for determination of the MPI cross-section as follows. Total electron numbers ($N_e$) generated by the femtosecond laser pulse are measured using Rayleigh Microwave Scattering (RMS) technique (see Methods and Supplementary Materials for details). Spatial and temporal intensity distribution $I(\mathbf{r}, t)$ are determined in precise measurements of laser beam and integral in the right-hand side is calculated. Then, one can determine $\sigma_m$ from the Eq. (2) for the known background gas density $n_0$.

General expression (2) can be simplified if additional assumptions are made. Firstly, we will consider in this work the most practical case of Gaussian beam. In this case spatial and temporal intensity dependences can be expressed as follows:

$$I(r,z,t) = I_0 \frac{w_0^2}{w(z)^2} e^{-\frac{2r^2}{w(z)^2}} e^{-\frac{(t-t^*)^2}{\tau^2}} \tag{3}$$

where $I_0$– intensity in the beam center, $w_0$ - $1/e^2$ waist radius (at $z=0$), $w(z) = w_0\sqrt{1 + \left(\frac{z}{z_R}\right)^2}$ – $1/e^2$ beam radius at location $z$ along the beam, $z_R$- Rayleigh length and $\tau$ – characteristic temporal width of the beam. This approximation uses standard for the non-dispersing medium Gaussian beam optics spatial distribution ($I_\mathbf{r}(r,z) = I_0 \frac{w_0^2}{w(z)^2} e^{-\frac{2r^2}{w(z)^2}}$) with Gaussian temporal shape $e^{-\frac{(t-t^*(r,z))^2}{\tau^2}}$, where $t^* = t^*(r,z)$ indicates moment of time when the beam reaches particular $(r,z)$-location (it is taken that beam peak reaches the waist at $z=0$ at time $t=0$, so that $t^*(r,0) = 0$ and $t^*(0,z) = z/c$)[26,27]. All parameters of the beam in Eq. (3), namely $I_0$, $w_0$, $z_R$ and $\tau$ are determined experimentally.

Secondly, we will consider in this work the case of atmospheric air and 800 nm laser. In this case, MPI of oxygen molecules is dominant process since O$_2$ photoionization rate is 2-3 order larger than that for N$_2$ due to its lower ionization potential[5,24]. Thus, using ionization energy of oxygen molecule $\mathcal{E}_i$=12.2 eV and energy of 800 nm ionizing photons of $\hbar\omega$ =1.55 eV, one can see that eight-photon photoionization process should be considered, namely $m$=8.

Simplified form of the expression (2) for MPI of air with femtosecond laser pulse of Gaussian shape in temporal and spatial domains can be now deduced by analytical integration of the intensity in the form (3), namely: $\int\int I(r,z,t)^8 dV dt = \frac{231\pi}{1024\cdot 16}\sqrt{\frac{\pi}{8}} I_0^8 \pi w_0^2 z_R \tau$ (see Supplementary Materials) and plugging it to the right-hand side of the Eq. (2). Finally, Eq. 2 can be reduced to the form:

$$N_e = \frac{231\pi}{1024 \cdot 16}\sqrt{\frac{\pi}{8}} \sigma_8 n_0 \tau \pi w_0^2 z_R \cdot I_0^8 \tag{4}$$

The determination of MPI cross-section of oxygen is conducted using Eq. (4) as follows. $N_e$ in the left-hand side of the equation is measured using Rayleigh Microwave Scattering (RMS) technique (see Methods and Supplementary Materials). Spatial and temporal characteristics of the laser beam ($I_0$, $w_0$, $z_R$ and $\tau$) in the right-hand side of the Eq. (4) are measured directly. Then, experimental dependence of $N_e$ vs. laser intensities $I_0$ is plotted and $\sigma_8$ is determined by obtaining the best fit of that dependence using Eq. (4).

**Experimental Details**

The experimental layout including femtosecond laser and RMS system are shown schematically in Figure 1. Photoionization of air (relative humidity 30%, temperature about 300 K) was achieved by focusing laser pulses having Gaussian temporal and spatial shape from a 800 nm Ti:Sapphire laser of 164 fs FWHM having repetition rate of 100 Hz using a 1000 mm plano-convex lens. The laser repetition rate was decreased from nominal 1 kHz to ensure no memory effect in the interrogated volume. Diameter of incident beam on the lens was 7 mm. The pulse energy was varied using a linear polarizer and measured using laser power meter (Gentec-EO XLP12-3S-H2-DO). Images of the plasma were taken using 1024i Pi Max 4 ICCD camera. Coordinate $z=0$ was chosen at the beam focal plane of the strongly attenuated laser beam (no plasma presents).

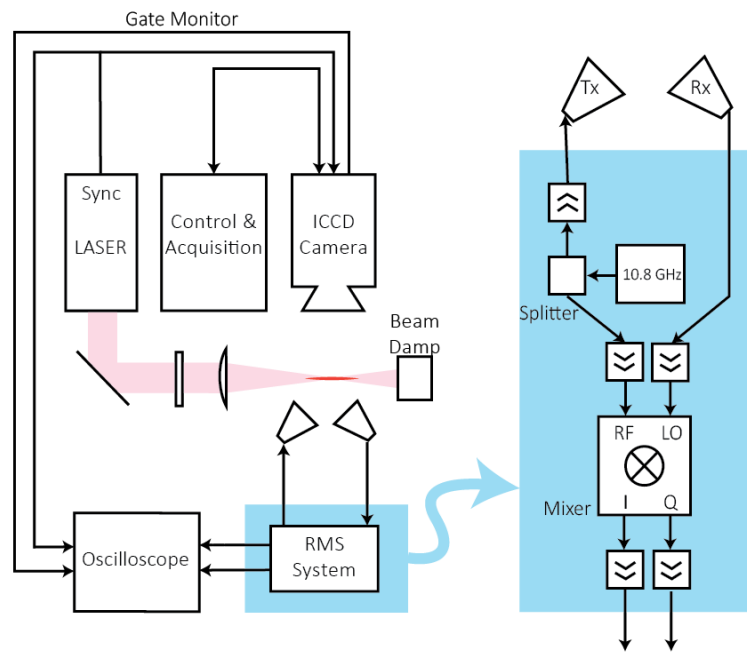

**Figure 1. Experimental Setup and RMS homodyne system.** A 800 nm Ti:Sapphire femtosecond laser operating at repetition rate of 100 Hz was focused using a 1000 mm plano-convex lens to create the plasma. Homodyne-type Rayleight microwave scattering system calibrated against dielectric scatterers was used to measure absolute number of electrons in the plasma volume.

Measurements of the total number of electrons in the plasma volume were conducted using Rayleigh Microwave Scattering (RMS) diagnostics (see Methods and Supplementary Materials). Homodyne-type RMS system operating at the microwave frequency 10.8 GHz was used as shown schematically in Figure 1. The microwave signal from the source was splitted in two arms: one arm sent microwaves to the plasma using radiating horn, while second arm delivered the signal directly to LO-input of the $I/Q$ mixer. Microwave radiation was linearly polarized along the plasma channel orientation. Radiating and detecting horns were mounted at the distance 6 cm from the plasma. The signal scattered from the plasma was received by the detecting horn, amplified and sent to the RF-input of $I/Q$ mixer. The two outputs of the $I/Q$ mixer were again amplified and captured on the oscilloscope. All components of the microwave system operated in the linear range of powers to ensure the measured response proportional to the amplitude of signal scattered from the plasma volume. The overall time response of the system was measured to be about 250 ps.

**Measurements of MPI cross-section of oxygen**

Spatial distribution of laser beam intensity was determined using beam profiler measurements conducted with a strongly attenuated laser beam. To this end, set of attenuator plates was used that reduced beam intensity manifold (about 2-3 orders of magnitude) to completely eliminate plasma creation. $1/e^2$ radius of the beam measured using beam profiler at various $z$-locations is shown in Figure 2 (a-b). Location of beam waist refers to coordinate $z=0$. Waist radius of in $x$- and $y$-directions were $R_x$=92.17 μm and $R_y$=94.41 μm respectively. Average radius of the beam ($w$) was chosen to satisfy condition $\pi w^2 = \pi R_x R_y$ at each measured $z$-locations and was approximated by analytical function $w(z) = w_0\sqrt{1 + \left(\frac{z}{z_R}\right)^2}$ with waist radius $w_0$=93.28 μm and Rayleigh length $z_R$=26.98 mm to achieve the best fit of the experimental data.

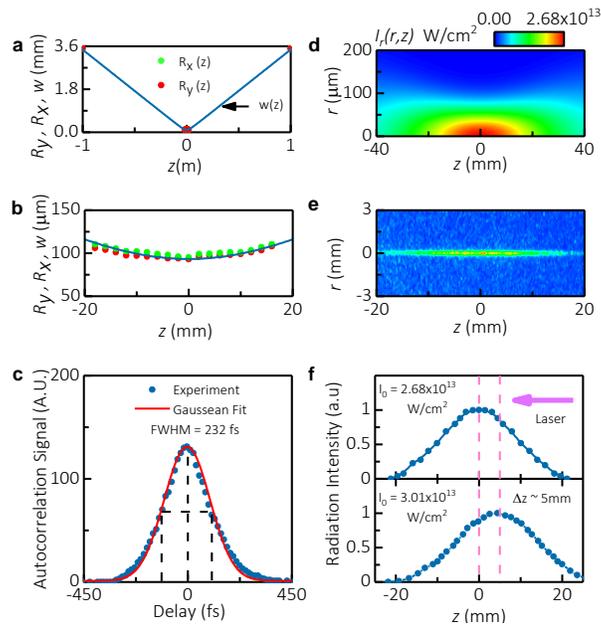

**Figure 2.** Measurements of the laser beam properties for intensity in the center $I_0$=2.68 x 10$^{13}$ W/cm². **a-b,** $1/e^2$ radius of the laser beam at various $z$-locations. **c,** Intensity autocorrelation function approximated by the Gaussian fit. **d,** 2D spatial distribution of the laser beam intensity approximated by Gaussian fit $I_r(r,z) = I_0 \frac{w_0^2}{w(z)^2} e^{-\frac{2r^2}{w(z)^2}}$ based on measured laser beam radius and Rayleigh length. **e,** Photographs of laser-induced plasma taken at exposure time of 10 ns by ICCD camera. **f,** Distribution of intensity radiated by the plasma plotted along the z-axis for two laser intensities.

Temporal shape of the laser pulse was determined using measurements of intensity autocorrelation function by means of 2$^{nd}$ harmonic generation crystal. The autocorrelation function had nearly Gaussian shape with full width at half maximum (FWHM) equal to $FWHM_\tau = 232$ fs as shown in Fig. 2(c). Thus, it may be concluded that laser intensity in the time domain was also Gaussian with $FWHM_t = \frac{232 \text{ fs}}{\sqrt{2}} =$ 164 fs. Finally, temporal dependence of the laser intensity was approximated by Gaussian distribution $I \propto e^{-\frac{t^2}{\tau^2}}$ with $\tau = \frac{FWHM_t}{2\sqrt{\ln 2}}$ = 98.6 fs.

Measured temporal and spatial parameters of the femtosecond laser pulse utilized in this work are summarized in the Table 1. Mean values averaged over the multiple measurements of the corresponding quantities and their standard errors are shown in the Table 1

| | Table 1\| Measured time-space parameters of the laser pulse | |
|---|---|---|
| $w_0$ | $1/e^2$ waist radius | 93.28±0.66µm |
| $z_R$ | Rayleigh length | 26.98±0.33mm |
| $\tau$ | Temporal width | 98.6±5.15fs |

Optical images of the laser-induced plasma created by MPI of air were analyzed to demonstrate that nonlinear effects in the non-attenuated laser beam were small and, thus, intensity approximation used in Eq. (3) still applies when the plasma was on. Typical photograph of the laser-induced plasma taken by ICCD camera (exposure time $t$=0-10 ns) is shown in Figure 2 (e) (energy in pulse 620 µJ, intensity at the beam center $I_0 = 2.68 \times 10^{13}$ W/cm$^2$). Figure 2(f) shows corresponding distribution of plasma radiation intensity ($S$) along $z$ –axis for two laser intensities $I_0 = 2.68 \times 10^{13}$ W/cm$^2$ and $3.01 \times 10^{13}$ W/cm$^2$. It was observed that focal plane of the beam coincided with coordinate $z$=0 for $I_0 \leq 2.68 \times 10^{13}$ W/cm$^2$. A shift of the focal plane toward the direction of the laser was observed for higher intensities, which can be explained by action of focusing Kerr nonlinearity. Based on that experimental evidence we have concluded that nonlinear effects (Kerr and plasma nonlinearities) were negligible for $I_0 \leq 2.68 \times 10^{13}$ W/cm$^2$.

Electron number generated by fs-laser laser pulse was measured using the RMS system shown in Figure 1. Figure 3(a) presents typical temporal evolution of number of electrons and amplitude of scattered microwave signal for two values of intensity $I_0 = 2.68 \times 10^{13}$ W/cm$^2$ and $2.93 \times 10^{13}$ W/cm$^2$. Right and left vertical axes indicate signal directly measured by the RMS and total number of electrons in the plasma volume $N_e$ determined using approach described in Methods. It was observed that plasma decayed faster for larger laser intensities, namely two-fold decay occurs on characteristic times 2.5 ns and 2 ns for $I_0 = 2.68 \times 10^{13}$ W/cm$^2$ and $2.93 \times 10^{13}$ W/cm$^2$ respectively.

Two distinct physical processes occurring on significantly different time scales can be traced on the Figure 3(a). First process is the fast rise at the moment of plasma creation (around $t$=0) associated with laser pulse passing the waist region and reaching the peak value of $N_e(t)$. Characteristic time of plasma creation is about the time required for light to pass the Raleigh length around the beam waist, namely $\frac{z_R}{c}$~0.1 ns. Second process is the decay of the plasma remaining after the laser pulse which occurs on characteristics times of about several nanoseconds according to the Figure 3(a) (see section below). RMS diagnostic was unable to temporally resolve the precise details of plasma creation due comparable response time of the system used (about 0.25 ns). However, RMS system precisely measured $N_e$ peak value and following plasma decay $N_e(t)$ since plasma recombination is occurring on significantly slower time scale, namely several nanoseconds and, therefore, difference between the true peak value and the measured value is negligible. Note, maximal electron number occurring immediately after the plasma creation is denoted as $N_e$ throughout the manuscript, while decay of the plasma refers to as the dependence $N_e(t)$. Figure 3(b) presents experimentally measured dependence of $N_e$ immediately after the plasma creation vs. intensity at the beam center $I_0$. This peak value $N_e$ is used for the purpose of determination of MPI cross-section $\sigma_8$.

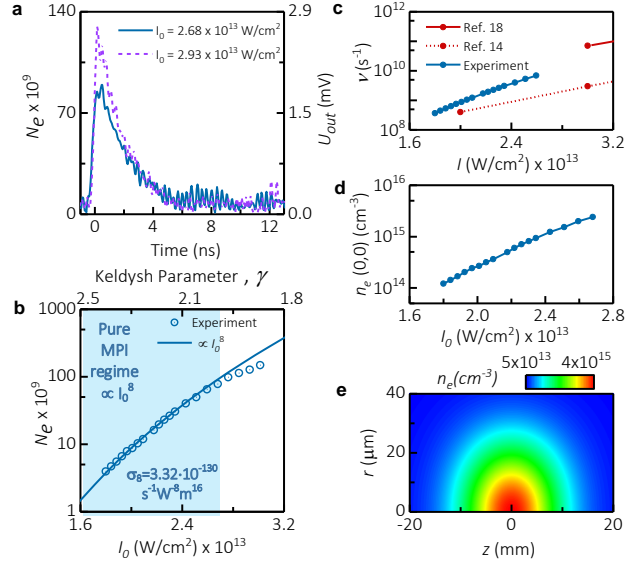

**Figure 3. Absolute measurements of parameters of MPI of air. a,** Temporal evolution of microwave signal scattered from the plasma $U_{out}(t)$ and total number of electrons in plasma volume $N_e(t)$. **b,** Measured dependence of $N_e$ immediately after the plasma creation vs. intensity at the beam center $I_0$. **c,** Comparison of theoretical and semi-empirical literature data with directly measured photoionization rate in this paper. Blue shaded area identifies range of intensities where pure MPI was observed $N_e \propto I_0^8$. **d,** Plasma density at the beam center $n_e(0,0)$ immediately after the laser pulse vs. intensity $I_0$. **e,** 2D distribution of plasma density immediately after the laser pulse for $I_0 = 2.68 \times 10^{13}$ W/cm².

Now we will determine MPI cross-section $\sigma_8$ by fitting the measured dependence of $N_e$ vs. $I_0$ shown in Figure 3 (b) using analytical expression (4). One can see that according to the analytical expression $N_e$ increases with laser intensity as $I_0^8$. RMS data shown in Figure 3(b) indicates that dependence $N_e \propto I_0^8$ was satisfied with high accuracy at low intensities $I_0 < 2.7 \cdot 10^{13}$ W/cm² which represents a clear manifestation of the pure MPI regime. Deviation from the $I_0^8$-law for higher intensities indicates departure from the pure MPI process at these higher $I_0$, which can be explained by relative proximity of Keldysh parameter $\gamma$ to 1 [top horizontal axis of the Figure 3(b)] and action of Kerr nonlinearity. Therefore, $N_e$ was fitted by the $I_0^8$-law for intensities $I_0 < 2.7 \cdot 10^{13}$ W/cm² and MPI cross-section $\sigma_8$ was determined based on the fit of this initial segment of the dependence as shown by the blue line in Figure 3(b) using parameters of the laser system measured above and density of molecular oxygen in the background air $n_0 \approx 5.13 \cdot 10^{18}$ cm⁻³. Finally, MPI cross-section was determined to be $\sigma_8 = (3.32 \pm 0.30) \times 10^{-130}$ W/cm².

**Photoionization rates, electron and species' densities**

Now we are going to consider oxygen photoionization rates based on the measured data and compare it with data available in literature. For laser beam center intensities $I_0 < 2.7 \cdot 10^{13}$ W/cm$^2$ (pure MPI regime), dependence of photoionization rate can be readily plotted as $\nu = \sigma_8 I^8$ shown by the solid blue curve in Figure 3(c) using value of $\sigma_8$ obtained above. Comparison with previously available data determined based on theoretical and semi-empirical approaches is also shown in Figure 3(c)[20,24]. One can see that semi-empirical predictions given in ref. 20 underestimated the photoionization rates about 2-3 times, while purely theoretical predictions in ref. 24 seems to slightly overestimate the rates.

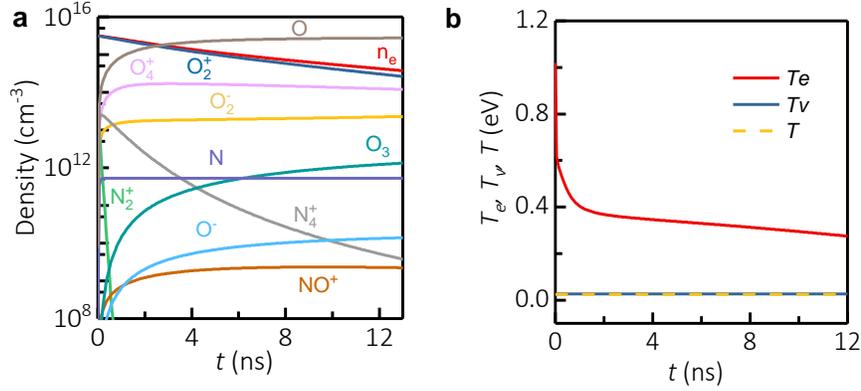

**Figure 4. Numerical simulations of plasma decay.** Plasma decay of after fs-laser pulse for $I_0 = 2.7 \cdot 10^{13}$ W/cm$^2$. **a**, Plasma species **b**, Electron, vibrational and gas temperatures.

Experiments conducted here paving the way to determine plasma density distribution created in the fs-laser induced plasmas. Distribution of the plasma density immediately after the laser pulse for the laser the Gaussian intensity distribution used in this work can be written using Eq. (1) as: $n_e(r,z) = \sigma_8 n_0 \int I(r,z,t)^8 dt = n_e(0,0) \frac{w_0^2}{w(z)^2} e^{-\frac{2r^2}{w(z)^2}}$. Integrating left and right side of this expression relates the plasma density at the origin location immediately after the laser pulse $n_e(0,0)$ with the directly measured quantities of $N_e$, $w_0$ and $z_R$ as follows: $n_e(0,0) = \frac{N_e}{\frac{231\pi}{1024 \cdot 16} \pi w_0^2 z_R}$. Figure 3(d) shows dependence of $n_e(0,0)$ on laser intensity. 2D distribution of plasma density $n_e(r,z)$ for $I_0 = 2.68 \times 10^{13}$ W/cm$^2$ is shown in Figure 3(e).

We have also numerically simulated the plasma decay to validate our experimental measurements. Plasma decay was simulated using 1D axially symmetric model in radial direction which self-consistently integrates Navier-Stokes, electron heat conduction, and electron-vibration energy transfer equations[25]. The model accounts for recombination of molecular ions, attachment processes, formation and decay of complex ions, electron energy losses due to electronic, vibrational excitations and elastic scattering.

Simulated decay of the densities and temperatures of various plasma species at the point of origin are shown in Figure 4 for intensity in the beam center $I_0 = 2.7 \times 10^{13}$ W/cm$^2$. Extremely fast (< 1ns) decrease of electron temperature to about 0.4 eV is associated with electron energy loss due to vibrational excitation of molecules, while slower later decay is governed by elastic collisions. Plasma density in the center decayed twice in about 3.2 ns, which is primarily dominated by dissociative and three-body recombination of molecular ions. Slightly faster plasma decay times observed in the experiments (about 2.5 ns) might be related to presence of water vapor in the ambient air.

**Concluding remarks**

In this work we have presented methodology which is paving the way for precise determination of the physical constants of multi-photon ionization, namely cross-section and ionization rate. The method utilizes precise measurement of the spatial and temporal distributions of the laser beam intensity and absolute measurement of total electron number in the plasma volume by means of elastic scattering of microwaves off the plasma volume and absolute calibration of the microwave system using dielectric scatterers. We have demonstrated capability of this method on the example of eight-photon ionization of molecular oxygen and determined the corresponding MPI cross-section to be $\sigma_8=(3.32\pm0.30)\cdot10^{-130}$ $W^{-8}m^{16}s^{-1}$. Future studies may be focused on precise tabulation of the cross-sections and photoionization rates of the multi-photon ionization for different gases and laser wavelengths using the methodology proposed and validated in this work. This effort would provide critical experimental evidence for the theoretical modeling of laser-induced plasmas.

**Acknowledgements**

We thank Dr. T. Meyers and Dr. D. Kartashov for useful discussions. This work was partially supported by NSF/DOE Partnership in the Basic Plasma Science and Engineering program (Grant No. 1465061).

# Appendix: Supplementary Information

### Rayleigh Microwave Scattering method description

To more clearly illustrate the method of absolute measurement of electron number in plasma volume we present the schematics of the Rayleigh Microwave Scattering (RMS) system in Figure 5 and provide more detailed description here. In Rayleigh Microwave Scattering technique, elastic scattering of microwave radiation off the plasma volume is measured and then total number of electron in the plasma volume is determined. The scattered radiation is created as result of polarization of the plasma channel in the external microwave field. For the thin plasma channel, when amplitude of the microwave field is nearly uniformly distributed inside the plasma, the radiation in far-field is equivalent to the Hertz dipole radiation. Overall, such a process is analogous to elastic Rayleigh scattering of light, when radiation wavelength significantly exceeds the scatterer size. Signal scattered from the plasma is proportional to the total electron numbers in the plasma volume. Absolute calibration of the RMS system was conducted using dielectric scatterers with known physical properties.

Linearly polarized microwave radiation at frequency of 10.8 GHz was scattered on the collinearly-oriented plasma filament and then amplitudes of the scattered signal was measured. Microwaves are radiated and detected using a horn as shown in Figure 5 mounted at the distance 6 cm from the plasma. A homodyne-type detection system was used for the scattered microwave signal measurements by means I/Q mixer.

We first demonstrate that amplitude of electric field induced inside the scatterer channel as result of irradiation with microwaves was uniform throughout the channel and equals to the field amplitude $E_0$ in the incident wave. For slender prolate plasma channel with length ($l$) significantly exceeding the diameter ($d$) used in this work. For conditions of this experiments plasma channel can be considered thin compared to skin depth (so that $f[GHz] \leq \frac{2.5}{\sigma[\Omega^{-1}cm^{-1}] \cdot d[mm]^2}$)[28]. In this case amplitude of electric field induced inside the scatterer with dielectric permittivity $\varepsilon$ and conductivity $\sigma$ can be written as $E = \frac{E_i}{\sqrt{(1+k(\varepsilon-1))^2 + \left(k\frac{\sigma}{\varepsilon_0 \omega}\right)^2}}$, where $k$ - depolarization factor governed by the channel geometry, $E_i$ – amplitude of incident microwave electric field at the channel location[29,30,31]. For conditions of experiments conducted here the depolarization factor $k$ is small due to large aspect ratio $AR=l/d>>1$:

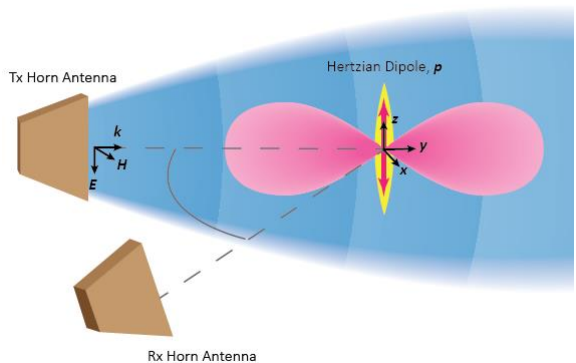

**Figure 5.** Detailed schematics of the Rayleigh Microwave Scattering (RMS)

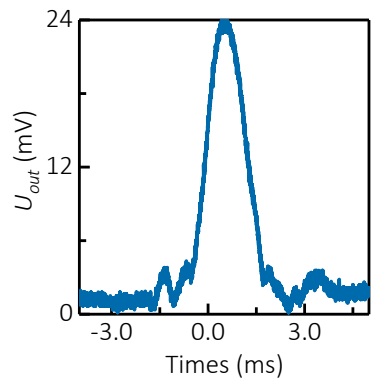

**Figure 6.** Calibration of the of the Rayleigh Microwave Scattering (RMS) system

$k \approx \frac{1}{AR^2} \ln(AR) \ll 1$ [29,30,31]. Therefore, the amplitude of electric field inside the channel is close to that in the incident wave $E = E_i$.[31,32,33] Flatness of the wave front surface along the scatterer length was ensured by placing the plasma scatterer at distance $r > \frac{l^2}{\lambda} \approx 1$ cm.

Electrons in the plasma volume experience oscillations with amplitude of $s = \frac{eE_i}{m\nu\omega}$ due to the incident microwave field (we consider no restoring force due prolate plasma channel geometries)[34]. Electron collision frequency in denominator is governed by those with gas particles ($\nu_{eg}$) for densities <$10^{17}$ cm$^{-3}$, since contribution of electron-ion collisions can be neglected in that range, so that finally $s = \frac{eE_i}{m\nu_{eg}\omega}$. Electron-gas collision frequency $\nu_{eg}$ is independent on plasma density ensuring that electrons in each location inside the plasma volume experience essentially same displacement. We used $\nu_{eg} = 5.18 \cdot 10^{11}$ s$^{-1}$ based on electron-gas elastic collision cross-section $\sigma_{eg} = 5 \cdot 10^{-16}$ cm$^{-2}$ and electron temperature $T_e$=0.4 eV.

Total dipole moment of the plasma channel ($p$) can be calculated now as follows:
$$p = es \int n(r,z) 2\pi r dr dz = esN_e = \frac{e^2}{m\nu\omega} \frac{E_i}{} N_e \qquad \text{Eq. (1)}$$

Radiation from the plasma dipole in plane perpendicular to the dipole orientation was detected by the same horn antenna as incident one (see Figure 5). The antenna was placed at distance $r$ =6 cm from the plasma channel to ensure that dominant contribution of far-field $\left(\sim \frac{k^2 p}{r}\right)$, while near-field $\left(\sim \frac{p}{r^3}\right)$ is negligible ($kr$ >6).[32,35] Thus, amplitude of electric field at location of the detecting horn was:
$$E_s = \frac{k^2 p}{r} = \frac{e^2}{mc^2\nu} \frac{\omega E_i}{r} N_e \qquad \text{Eq. (2)}$$

A homodyne-type detection system was used for the scattered microwave signal measurements which provides output voltage $U_{out} \propto E_s$. The detection was achieved by means of *I/Q* Mixer, providing in-phase (*I*) and quadrature (*Q*) outputs. The total amplitude of the scattered microwave signal is determined as: $U_{out} = \sqrt{I^2 + Q^2}$. The amplifiers and the mixer used in the microwave system are operating in a linear mode for the entire range of the scattered signal amplitudes, thereby ensuring that the output signal $U_{out}$ is proportional to the electric field amplitude of scattered radiation $E_s$ at the detection horn location: $U_{out} \propto E_s$.
$$U_{out} \propto E_s = \frac{e^2}{mc^2\nu} \frac{\omega E_i}{r} N_e \qquad \text{Eq. (3)}$$

One can see that RMS system detects total number of electrons in plasma volume accurate to coefficient function of specific microwave system used. Absolute calibration of the RMS can be conducted using dielectric scatterers with known physical properties. To this end, now we consider RMS system signal generated by the prolate scatterer made of dielectric material with dielectric constant $\varepsilon$ and volume $V$. The only difference from above consideration for the plasma channel would be that total dipole moment induced in the scatterer is $p = \varepsilon_0(\varepsilon - 1)E_i V$, and thus:
$$U_{out} \propto E_s = \frac{\varepsilon_0(\varepsilon-1)}{c^2} \frac{\omega^2 E_i}{r} V \qquad \text{Eq. (4)}$$

One particularly convenient form of expression for measured output of RMS system is:

$$U_{out} = \begin{cases} A \cdot \frac{e^2}{m\nu} N_e & - \text{ for plasma} \\ A \cdot V \cdot \varepsilon_0 (\varepsilon - 1)\omega & - \text{ for dielectric bullet} \end{cases} \quad \text{Eq. (5)}$$

where $A$ – proportionality coefficient which is a property of the specific microwave system (utilized components, geometry, microwave power, etc.) while independent of scatterer properties and it can be found using scatterers with known properties. Lower Eq. 5 was used for calibration of specific microwave system, namely in order to determine value of coefficient $A$. Cylindrical dielectric bullets with diameter 3.175 mm and length 1 cm made of Teflon were used. The bullets were shot through the microwave field (along the same axis where plasma was placed later) using pneumatic gun with velocities below 100 m/s in order to generate time-varying response on the dielectric bullet passage for separation of signal scattered from the bullet from one DC background caused by reflections from surroundings, elements of microwave circuit etc.

Calibration of the RMS system with 1 cm long and 3.175 mm diameter Teflon bullet yielded response of the RMS system shown in Figure 6. Coefficient A was found to be A=2.121·10⁵ V·Ω·m⁻² in this calibration procedure. Thus, relation between the total electron number in the plasma volume $N_e$ and amplitude of scattered system measured using RMS system $U_{out}$ was: $N_e = U_{out} \cdot 4.81 \cdot 10^{13}$.

**Calculating of intensity integral**

We present calculation of laser intensity integral over spatial and temporal variables:

$$\int \int I(r,z,t)^8 dt dV = \frac{231\pi}{1024 \cdot 16} \sqrt{\frac{\pi}{8}} I_0^8 \pi w_0^2 z_R \tau$$

The integration is first taken over time period when laser pulse exists at particular location of space and then spatial integral is taken over region where laser beam presents. For the specific system utilized in current experiment, main contribution to the temporal integral is gained during the laser pulse time of $\tau$=98.6 fs, while spatial integral is accumulated primarily in vicinity of the beam waist where intensity is maximal.

The limits of the integrals can be extended to infinity when analytical approximation space-time dependence of the laser pulse in the form $I(r,z,t) = I_0 \frac{w_0^2}{w(z)^2} e^{-\frac{2r^2}{w(z)^2}} e^{-\frac{(t-t^*(r,z))^2}{\tau^2}}$ with $w(z) = w_0 \sqrt{1 + \left(\frac{z}{z_R}\right)^2}$ is used:

$$\int \int I(r,z,t)^8 dt dV = \int_{-\infty}^{+\infty} \int_0^{+\infty} \int_{-\infty}^{+\infty} I_0^8 \frac{w_0^{16}}{w(z)^{16}} e^{-\frac{16r^2}{w(z)^2}} e^{-\frac{8(t-t^*(r,z))^2}{\tau^2}} 2\pi r \, dt \, dr \, dz$$

Inner temporal integral can be calculated analytically by changing of variables to $\tilde{t} = \frac{\sqrt{8}(t-t^*(r,z))}{\tau}$ for any finite $(r,z)$-location $\int_{-\infty}^{+\infty} e^{-\frac{8(t-t^*(r,z))}{\tau^2}} dt = \frac{\tau}{\sqrt{8}} \int_{-\infty}^{+\infty} e^{-\tilde{t}^2} d\tilde{t} = \sqrt{\frac{\pi}{8}} \tau$. The outer spatial integral can be first simply calculated over $r$ – variable and then using standard integral $\int_{-\infty}^{+\infty} \frac{1}{[1+x^2]^7} dx = \frac{231\pi}{1024}$. This finally yields:

$$\int\int I(r,z,t)^8 dt dV = I_0^8 \sqrt{\frac{\pi}{8}}\tau \int_{-\infty}^{+\infty}\int_0^{+\infty} \frac{w_0^{16}}{w(z)^{16}} e^{-\frac{16r^2}{w(z)^2}} 2\pi r dr dz = I_0^8 \sqrt{\frac{\pi}{8}}\tau \int_0^{+\infty} \frac{\pi w_0^{16}}{16} \frac{1}{w(z)^{14}} dz =$$

$$I_0^8 \sqrt{\frac{\pi}{8}}\tau \frac{\pi w_0^2}{16} \int_{-\infty}^{+\infty} \frac{1}{\left[1+\left(\frac{z}{z_R}\right)^2\right]^7} dz = I_0^8 \sqrt{\frac{\pi}{8}}\tau \frac{\pi w_0^2}{16} \frac{231\pi}{1024} z_R$$